\documentclass[pra,aps,showpacs,floatfix,notitlepage,reprint,nofootinbib]{revtex4-2}
\usepackage{amsmath}
\usepackage{graphicx}
\usepackage[ansinew]{inputenc}
\usepackage{array}
\usepackage{color}

\usepackage{amsxtra}
\usepackage{amstext}
\usepackage{amssymb}
\usepackage{latexsym}
\usepackage{dsfont}
\usepackage{lipsum}
\usepackage[colorlinks=true,allcolors=blue]{hyperref}
\usepackage{ulem}
\usepackage{orcidlink}

\renewcommand{\Re}{\text{Re}}
\renewcommand{\Im}{\text{Im}}

\newcommand{\mb}[1]{\mathbf{#1}}
\newcommand{\bs}[1]{\boldsymbol{#1}}
\newcommand{\tm}[1]{\text{#1}}
\newcommand{\lel}{\left}
\newcommand{\rer}{\right}

\newcommand{\brakk}[3]{\left\langle{#1}\left|{#2}\right|{#3}\right\rangle}

\newcommand{\avgg}[1]{\left\langle\left\langle #1\right\rangle\right\rangle}

\newcommand{\comm}[2]{\left[#1,#2\right]}

\newcommand{\cD}{\mathcal{D}}

\newcommand{\cH}{\mathcal{H}}

\newcommand{\bk}{\mathbf{k}}

\newcommand{\cP}{\mathcal{P}}
\newcommand{\br}{\mathbf{r}}
\newcommand{\rb}{\mb{r}}

\newcommand{\cdd}{\cdot}
\newcommand{\la}{\lambda}
\newcommand{\om}{\omega}

\newcommand{\da}{\dagger}

\newcommand{\thh}{\theta}
\newcommand{\st}{\star}

\newcommand{\curl}[1]{\boldsymbol{\nabla}_{#1}\times}

\newcommand{\dive}[1]{\boldsymbol{\nabla}_{#1}\cdot}

\def\XXint#1#2#3{{\setbox0=\hbox{$#1{#2#3}{\int}$}
     \vcenter{\hbox{$#2#3$}}\kern-.5\wd0}}

\newcommand{\intkf}{\int \frac{d^3k}{(2\pi)^3} \;}

\newcommand{\intw}{\int_{0}^{\infty} d\om\;}

\newcommand{\ints}{\int_{-\infty}^{\infty} ds\;}
\newcommand{\intsp}{\int_{-\infty}^{\infty} ds'\;}

\newcommand{\intr}{\int d^3r \;}
\newcommand{\intrp}{\int d^3r' \;}

\newcommand{\id}{\mathbb{I}}

\newcommand{\Eh}{\widehat{\mathbf{E}}}
\newcommand{\Dh}{\widehat{\mathbf{D}}}
\newcommand{\Hh}{\widehat{\mathbf{H}}}
\newcommand{\Bh}{\widehat{\mathbf{B}}}
\newcommand{\Sh}{\widehat{\mathbf{S}}}
\newcommand{\Ph}{\widehat{\mathbf{P}}}
\newcommand{\Mh}{\widehat{\mathbf{M}}}
\newcommand{\mhh}{\widehat{\mathbf{m}}}
\newcommand{\mh}{\widehat{\mathbf{m}}}
\newcommand{\dhh}{\widehat{\mathbf{d}}}
\newcommand{\fh}{\widehat{\mathbf{f}}}
\newcommand{\jh}{\widehat{\mathbf{j}}}
\newcommand{\rhoh}{\widehat{\rho}}

\newcommand{\gG}{\mb{\underline{G}}}

\begin{document}
\title{Duality, decay rates and local-field models in macroscopic QED}
\author{Niclas Westerberg \orcidlink{0000-0002-9876-5858} $^\da$}
\email{Niclas.Westerberg@glasgow.ac.uk}
\author{Anette Messinger \orcidlink{0000-0003-3208-3974} $^\da$}
\author{Stephen M. Barnett \orcidlink{0000-0003-0733-4524} }
\affiliation{School of Physics and Astronomy, University of Glasgow, Glasgow, G12 8QQ, United Kingdom}
\begin{abstract}
Any treatment of magnetic interactions between atoms, molecules and optical media must start at the form of the interaction energy. This forms the base on which predictions about any number of magnetic atom-light properties stands -- spontaneous decay rates and forces included. As is well-known, the Heaviside-Larmor duality symmetry of Maxwell's equations, where electric and magnetic quantities are exchanges, is broken by the usual form of the magnetic interaction energy. We argue that this symmetry can be restored by including general local-field effects, and that local fields should be treated as a necessity for correctly translating between the microscopic world of the dipole and the macroscopic world of the measured fields. This may additionally aid in resolving a long standing debate over the form of the force on a dipole in a medium. Finally, we compute the magnetic dipole decay rate in a magneto-dielectric with local-field effects taken into account, and show that macroscopic quantum electrodynamics can be made to be dual symmetric at an operator level, instead of only for expectation values. 
\end{abstract}

\date{\today}
\maketitle

\noindent 

\section{Introduction}\label{sec:intro}
At the heart of both classical \cite{stratton, jackson} and quantum electrodynamics \cite{cohen, loudon} lies the coupling between light and matter, and all our predictions for emission rates, atomic scattering and forces \cite{dispersion}, to name a few, depend crucially on this. Indeed, along with the advance of quantum technology, we rely on ever more precise predictions of the interaction between light and both atoms and artificial qubits \cite{qt1, qt2, qt3}. In a wide variety of optical media, theory and experiments are in excellent agreement \cite{CasimirReview,ultrastrong}. The issue lies with emitters embedded in magnetic media, such as magneto-dielectric or magneto-electric media, where the magnetic response plays an important role. As is well-known \cite{stratton,jackson}, naturally occurring optical media with a significant magnetic response are very rare. However, with the advent of metamaterials where a magnetic response can be engineered \cite{metamaterials1, metamaterials2}, this is becoming increasingly important, often arising in conjunction with anisotropic dielectric responses \cite{metamaterialsReview, anetteMeSteve}. 

The form of the magnetic light-matter interaction energy is closely related to the force on a dipole inside a medium, the exact form of which has been under much scrutiny and debate \cite{forces1, forces2, stevesForces, forces3, forces4, forces5}. The question can, in essence, be summarised by whether the interaction energy of a magnetic dipole $\mhh$ scales with the magnetic field $\Hh$ or the magnetic induction $\Bh = \mu\Hh$. Such a magnetic dipole can be either the exact point-dipole of a spin, or relate to the angular momentum properties of an electric dipole. The question is then, how does a magnetic dipole couple to the electromagnetic field? In vacuum and (most) natural media, it is clear that this is a non-issue, since $\mu = \mu_0$ in such cases. We will note that there is an interesting discrepancy between the literature of macroscopic quantum electrodynamics and quantum optics, and that which treats dia-, para- and ferromagnetism. In the former, we find an interaction energy of the form $\cH_\tm{int} =-\mhh\cdd\Bh$ \cite{dispersion, molecularQEDbook}, whereas in the latter the form $\cH_\tm{int} =-\mu_0\mhh\cdd\Hh$ is favoured \cite{magnets}, which notably is in agreement also with Ref.~\cite{einstein1908}. In the magnetic literature, importance of the local field is also stressed. We will return to this point, and also note that due to the quantum mechanical nature of magnetism (i.e. spin), it can be expected that a derivation through a classical (point-like) Lagrangian may not be appropriate. Nonetheless, the discussion here is not on which field is more `fundamental', it is simply about how to correctly link the world of microscopic fields to those of the macroscopic world.

This manuscript is organised as follows: in Section~\ref{sec:dual}, we will present arguments for why a dual symmetric coupling is desirable, which is followed by a discussion of the mathematical framework in Section~\ref{sec:formulation}. We then present some results in Sections~\ref{sec:decay} and \ref{sec:LL}, where we compute the spontaneous decay rate of a dipole in magnetic media: first in the absence of local-field corrections, which include a correction for absorbing media as compared to the standard result, which is then followed by a dual symmetric result. We finish the manuscript with concluding remarks in Section~\ref{sec:conc}.

\section{An argument for a dual symmetric coupling}\label{sec:dual}

\begin{figure}
    \centering
    \includegraphics[width=0.35\textwidth]{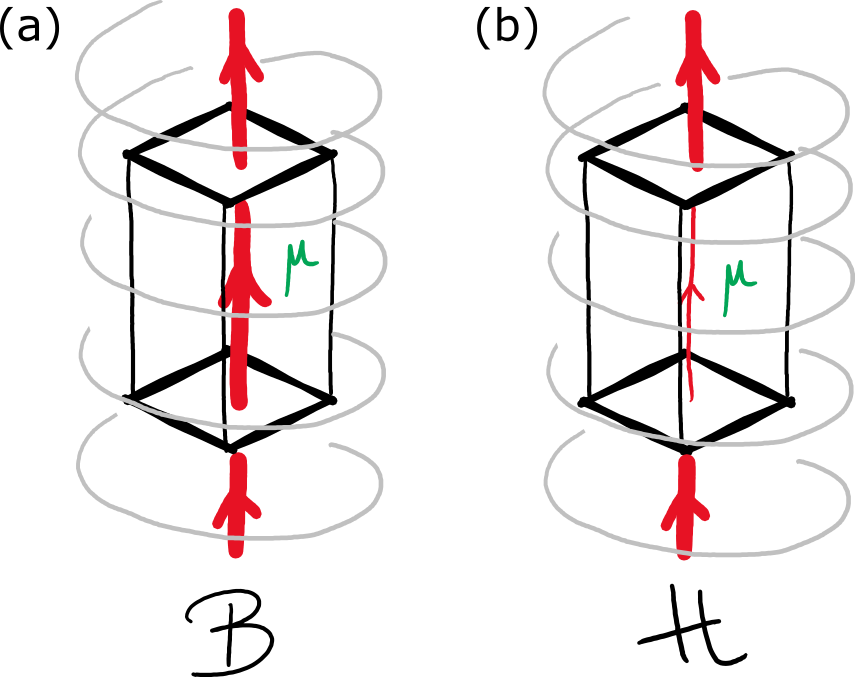}
    \caption{\textbf{(a)} The magnetic induction $\Bh$ from a solenoid is unchanged by the presence of a paramagnetic medium. \textbf{(b)} The corresponding magnetic field $\Hh$, which is screened by the paramagnetic medium, represented by a thinner line.}
    \label{fig:sketch}
\end{figure}

Let us start this discussion by appealing to a physical situation: suppose we embed a magnetic dipole inside a magnetic medium, which sits inside a solenoid such that a constant field is applied. A sketch of this can be seen in Fig.~\ref{fig:sketch}. By applying the boundary condition across the medium surface \cite{stratton, jackson}, we find that the magnetic induction is continuous and thus an interaction energy of the form $\cH_\tm{int} =-\mhh\cdd\Bh$ is unchanged by the presence of the medium. We find this rather surprising. Indeed, we would expect the interaction energy to change, similarly to the electric case. This indicates that something more is play, and lies at the heart of the argument which we appeal to here.

To put it in context, let us consider an electric dipole. Typically, the electric dipole $\dhh$ is a useful construction when the wavelength of light $\la$ is much larger than the typical length-scale of the electronic motion. This is the case both classically \cite{jackson} and at a quantum level \cite{cohen}, and depending on the context, the interaction energy is written in the form $\cH_\tm{int} =- \dhh\cdd\Eh$ or $\cH_\tm{int} =-\varepsilon_0^{-1}\dhh\cdd\Dh$, where $\Eh$ and $\Dh$ is the electric and displacement field operators respectively. The former form is usually favoured semi-classically as well as in the context of macroscopic QED \cite{scheel, dispersion}, whereas the latter is more common in cavity and molecular QED \cite{molecularQEDbook}. This has naturally caused some confusion over which field to use, resolved by \citet{milonni}, although it has once more come to the forefront of research as this choice (commonly controlled by a gauge choice \cite{cohen} but not necessarily \cite{Woolley}) re-distributes the energy between matter and field components -- something that can lead to gauge ambiguous predictions if energy-dependent approximations are made further along the calculation \cite{adam1, nori1, nori2, myself, adam2}. Regardless, in the context of macroscopic QED, we would like to note that the `electric' field usually \cite{dispersion} referred to when writing $\cH_\tm{int} = -\dhh\cdd\Eh$ is, in fact, a type of displacement field. It is not the full displacement field accounting for all bound charges (which would include the dielectric medium), but a more careful consideration \cite{theChemPaper} shows that the $\Eh$-operator contains the polarisation induced by the dipole, similarly to cavity QED. This is important for conceptual reasons here, as we want to stress that there is some sense of choice present here. In calculations however, this displacement field $\Eh$ behaves as the classical electric field without the presence of the electric dipole. As can be seen, the form of the coupling is not as clear-cut as one would first believe.

Formally, the discussion around Fig.~\ref{fig:sketch} revolves around the Heaviside-Larmor symmetry of Maxwell's equations. This is a type of duality symmetry of Maxwell's equations where electric and magnetic quantities can be exchanged without altering the dynamics, as was first discussed by Heaviside \cite{heaviside} and Larmor \cite{larmor}.\footnote{We should note that this is difficult to find in their writing.} For later reference, we will use the terminology `duality symmetry' interchangeably with `Heaviside-Larmor symmetry' due to the ubiquity of both terminologies. Of import here, Heaviside-Larmor symmetry holds true for both the field energy density $\mathcal{H}_\tm{field}=\left[\Dh\cdd\Eh+\Bh\cdd\Hh\right]/2$ and Poynting vector $\Sh=\Eh\times\Hh$. In general, symmetry is a well-known and powerful guiding mechanism in theoretical physics, and an important point is that for a symmetry (of well-established equations) not to hold, there has to be some physical mechanism behind it. Here we find no such mechanism. We would like to stress that, akin to any other symmetry such as Lorentz symmetry, a duality transform does not alter the physics, only our description of it. Before we continue we must quickly introduce Maxwell's equations and the setting before we return to the Heaviside-Larmor symmetry. 

\section{Maxwell's equations, equivalent formulations and Heaviside-Larmor symmetry}\label{sec:formulation}

\begin{figure*}
    \centering
    \includegraphics[width=0.95\textwidth]{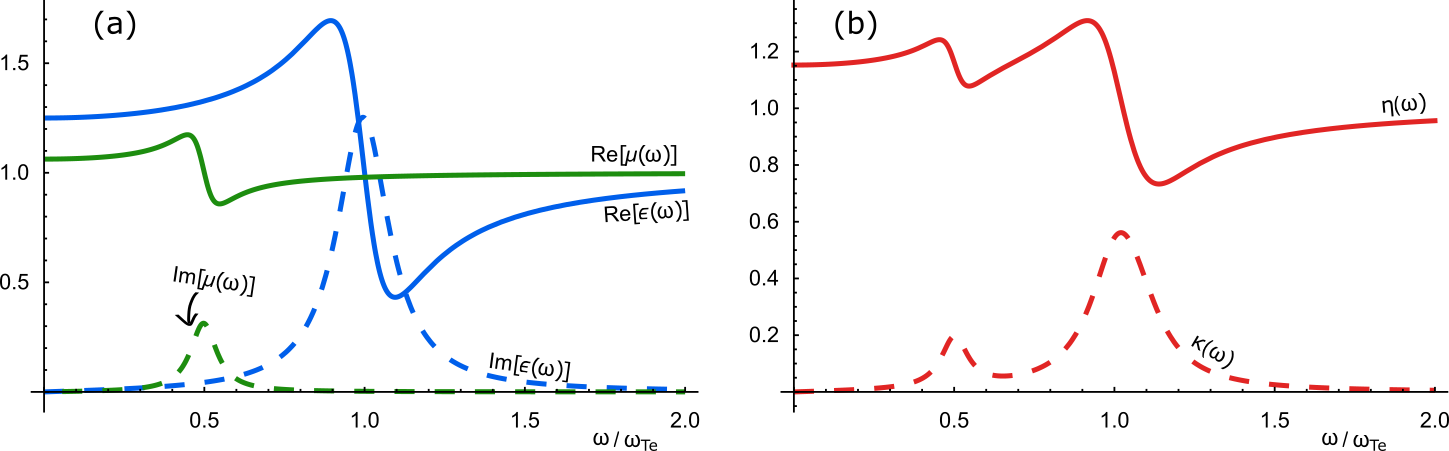}
    \caption{\textbf{(a)} Relative permittivity and permeability described by Eqns.~\eqref{eq:permitt} and \eqref{eq:permea} \textbf{(b)} The corresponding complex index of refraction $n(\om) = \sqrt{\varepsilon(\om)\mu(\om)} = \eta(\om) + i \kappa(\om)$.}
    \label{fig:disp}
\end{figure*}

We should first note that the points that we are about to make does not rely on a particular formulation of macroscopic QED, although we will specify an example later. For notational simplicity, we will from now on work in units such that $c = 1 = \varepsilon_0 = \hbar$. We will here consider a homogeneous magneto-dielectric (with permittivity $\varepsilon$ and permeability $\mu$) with an embedded quantum emitter of magnetic dipole moment operator $\mh$. Such a medium can also be described by the refractive index $n(\om) = \sqrt{\varepsilon(\om)\mu(\om)}$. Let us simply start at Maxwell's equations as usually written when discussing macroscopic QED:
\begin{align}
\dive{}\Dh &= 0,\label{eq:M1}\\
\dive{}\Bh &= 0,\label{eq:M2}\\
\curl{}\Eh &= -\partial_t \Bh,\label{eq:M3}\\
\curl{}\Hh &= \partial_t \Dh,\label{eq:M4}
\end{align}
where $\Eh$ is the electric field operator, $\Hh$ is the magnetic field operator, and we define the displacement field $\Dh = \Eh + \Ph_\tm{tot}$ and the magnetic induction $\Bh = \Hh + \Mh_\tm{tot}$ with $\Ph_\tm{tot}$ and $\Mh_\tm{tot}$ being the total polarisation and magnetisation field operators, respectively. Here we write the total polarisation $\Ph_\tm{tot}$ as the sum of the induced polarisation $\Ph(\Eh) = (\varepsilon-1)\Eh$ and the associated noise polarisation $\Ph_N$ required for absorption \cite{steves,scheel}: $\Ph_\tm{tot} = \Ph(\Eh)+\Ph_N$. 

Similar holds true for the magnetisation, although here convention is not as clear and we can make the choice of writing $\Mh(\Hh) = \left(\mu-1\right)\Hh$ or $\Mh(\Bh)=\left(1-\mu^{-1}\right)\Bh$. This is, of course, related to our issue at hand. When we make this choice, convention dictates that we write the noise magnetisation $\Mh_N$ in two separate ways: 
\begin{enumerate}
    \item \label{option1} $\Mh_\tm{tot} = \Mh(\Hh)+\Mh_N +\mh$. 
    \item \label{option2} $\Mh_\tm{tot} = \Mh(\Bh)+\mu\Mh_N +\mh$. 
\end{enumerate}
This is only a convention that follows from whether we naturally associate the absorption, and therefore the noise, with $\mu$ or $\kappa = 1/\mu$. Although we have to note that we are free to also swap the convention, as it is purely a definition, we will for notational clarity label the different definitions for noise magnetisations $\Mh_{N,H}$ and $\Mh_{N,B}$, respectively. We will here simply demand that these two formulations are equivalent, and yield the same physical predictions, as is the case in the classical formulation. Option~\eqref{option2} is considered in Ref.~\cite{scheelDuality}, and we will therefore focus on option~\eqref{option1} and we show in Appendix~\ref{app:decayQuant} that our formulation here is indeed equivalent.

The quantisation can be done in an entirely analogues manner to for instance Ref.~\cite{stevesLocalField} by introducing the polariton operators $\fh_{\lambda}$ where the subscript $\lambda = \{e,m\}$ denote electric and magnetic quantities, respectively, and they obey the commutation relation
\begin{align}
\left[\fh_{\lambda,i}(\br,\om),\fh^\da_{\lambda',j}(\br',\om')\right]=\delta_{ij}\delta_{\lambda\lambda'}\delta(\br-\br')\delta(\om-\om').
\end{align}
It can further be shown that the field Hamiltonian is simply $H_\tm{field} = \sum_{\lambda=e,m}\intr \int_0^\infty d\om \; \om\,\fh^\da_\la(\br,\om)\cdd\fh_\la(\br,\om)$, whether this is approached using Fano diagonolisation \cite{steves, iranians2006,iranians2007, iranians2008} or Langevin noise \cite{scheelUnified, scheelAnother,scheelDuality, Matloob}. We refer to the previously mentioned references for a detailed discussion. In order to evaluate these expectation values, we will need to use the noise operators associated with the medium absorption. The fluctuation-dissipation theorem \cite{kubo,steves,stevesLocalField, dispersion} demands that 
\begin{align}\label{eq:flucDissP}
&\brakk{0}{\Ph_{N,i}(\mb{r},\om)\Ph^\da_{N,j}(\mb{r'},\om')}{0} \nonumber\\ 
&\quad\quad = \left(\frac{\Im[\varepsilon]}{\pi}\right)\delta_{ij}\delta(\mb{r}-\mb{r}')\delta(\om-\om'),\\
&\brakk{0}{\Mh_{N,H,i}(\mb{r},\om)\Mh^\da_{N,H,j}(\mb{r'},\om')}{0} \nonumber\\ 
&\quad\quad = \left(\frac{\Im[\mu]}{\pi}\right)\delta_{ij}\delta(\mb{r}-\mb{r}')\delta(\om-\om').\label{eq:flucDissM}
\end{align}
We can further relate the noise polarisation/magnetisation to the polaritons of the medium through
\begin{align}\label{eq:Hnoise}
\left(\begin{array}{c}
\Ph_N \\
\Mh_{N,H}
\end{array}\right) = \frac{1}{\sqrt{\pi}}\left(\begin{array}{cc}
i\sqrt{\Im{\;\varepsilon}}& 0 \\
0 & i\sqrt{\Im{\;\mu}}
\end{array}\right) \left(\begin{array}{c}
\fh_e \\
\fh_m
\end{array}\right).
\end{align}
Importantly, from Maxwell's equations \eqref{eq:M1}-\eqref{eq:M4}, we find that the macroscopic magnetic field can be written as
\begin{align}\label{eq:HinNoise}
\Hh(\rb,\om)= \intrp \gG^H(\rb,\rb',\om)&\cdd\bigg[\om^2\Mh_{N,H}(\rb',\om)\\
&-i\om\varepsilon^{-1}\curl{\rb'}\Ph_N(\rb',\om)\bigg]\nonumber
\end{align}
with the Green's function for $\Hh$ satisfying
\begin{align}\label{eq:GreensH}
\varepsilon^{-1}\curl{}\left[\curl{}\gG^H\right]-\om^2\mu \gG^H= \id \delta(\mb{r}-\mb{r}').
\end{align}
In other words, the magnetic Green's function $\gG^H$ and the electric Green's function $\gG^E$ reported in Ref.~\cite{stevesLocalField} (amongst others) is dual symmetric. 

As is well-known, Maxwell's equations \eqref{eq:M1}-\eqref{eq:M4} possesses Heaviside-Larmor symmetry, meaning that the duality transformation
\begin{align}\label{eq:dualityTransform}
\left(\begin{array}{c}
\Eh \\
\Hh
\end{array}\right) &\rightarrow \left(\begin{array}{cc}
\cos\thh & \sin\thh \\
-\sin\thh & \cos\thh
\end{array}\right) \left(\begin{array}{c}
\Eh \\
\Hh
\end{array}\right) = \left(\begin{array}{c}
\Eh^\star \\
\Hh^\star
\end{array}\right),
\end{align}
along with the same rotation for $\left(\Dh,\Bh\right)^T$, leaves Maxwell's equations unchanged, for any value of $\theta$. This implicitly leads to a similar transformation for the polarisation and magnetisation fields. We will here follow the notation of Ref.~\cite{scheelDuality} and refer to the rotation matrix in Eq.~\eqref{eq:dualityTransform} as $\cD(\theta)$. As is done in Ref.~\cite{scheelDuality}, we can use these constituent equations to relate the dual permittivities. We will here focus on a dual transformation with $\thh=\pi/2$ where $\varepsilon^\st = \mu$ and $\mu^\st = \varepsilon$, as it is sufficient to illustrate our point. Additionally, we should here note that the constituent equations for $\Dh$ and $\Bh$ also imply duality transformations for the noise polarisation/magnetisation. The exact transformation depends on whether we consider option~\eqref{option1} or \eqref{option2} as mentioned above. In particular, we find simply that
\begin{align}\label{eq:polDual}
\left(\begin{array}{c}
\Ph_N^\star \\
\Mh_{N,H}^\star
\end{array}\right) &= 
\left(\begin{array}{cc}
\cos\thh & \sin\thh \\
-\sin\thh & \cos\thh
\end{array}\right) \left(\begin{array}{c}
\Ph_N\\
\Mh_{N,H}
\end{array}\right) \nonumber\\
&\overset{\thh\rightarrow\pi/2}{=} \left(\begin{array}{c}
\Mh_{N,H}\\
-\Ph_{N}
\end{array}\right),
\end{align}
whereas the equivalent for option \eqref{option2} can be seen in Eq.~(8) of Ref.~\cite{scheelDuality}. Exactly the same relations hold for the polariton operators in this formulation. Nonetheless, we want to stress that there is no physical reason for the medium under consideration to break this Heaviside-Larmor symmetry. For a general $\thh$, we would naturally have to introduce a magneto-electric response where $\Ph(\Eh,\Hh)$ and $\Mh(\Hh,\Eh)$, since a duality transformation mixes electric and magnetic quantities, and it is not possible to pack this into a combination of $\varepsilon^\st$ and $\mu^\st$ only. A similar point is made in Ref.~\cite{scheelDualityExtended}, although the interpretation differ. Regardless, we stress that these transformations only affect our description of the physics, and physical predictions must remain the same: it reflects that relative permittivities and permeabilities are not unique quantities. For illustrative purposes, we will consider an example medium with 
\begin{align}\label{eq:permitt}
\varepsilon(\om) &= 1-\frac{\om_\tm{Le}^2}{\om^2-\om_\tm{Te}^2+2i\gamma_\tm{e} \om},\\
\mu(\om) &= 1-\frac{\om_\tm{Lm}^2}{\om^2-\om_\tm{Tm}^2+2i\gamma_\tm{m} \om}. \label{eq:permea}
\end{align}
An example of such a medium as can be seen in Fig.~\ref{fig:disp}, where we consider $\om_\tm{Le} = \om_\tm{Te}/2$, $\om_\tm{Le} = \om_\tm{Te}/2$, $\gamma_\tm{e} = \om_\tm{Te}/10$, $\om_\tm{Lm} = \om_\tm{Te}/8$ and $\om_\tm{Tm} = \om_\tm{Te}/2$. Note however that the considerations here are not limited to a medium of this form.

Let us return to the form of the interaction energy again, and start with the form most commonly found in the literature treating atoms embedded in some macroscopic media \cite{dispersion}: $\cH_\tm{int} = -\mhh\cdd\Bh$. If we apply a $\pi/2$-duality transform in Eq.~\eqref{eq:dualityTransform} (along with corresponding transforms for $\Bh$ and $\mu$), we find that the interaction energy in the dual representation is 
\begin{align}
\cH_\tm{int} = -\mhh^\st\cdd\Bh^\st = -\dhh\cdd\Dh \neq  -\dhh\cdd\Eh.
\end{align} 
This light-matter interaction is at odds with Heaviside-Larmor symmetry, and therefore at odd with expectation. There is thus an issue: an interaction energy of the form $-\mhh\cdd\Hh$ is dual symmetric, but in the atomic Hamiltonian we commonly find $-\mhh\cdd\Bh$, whose introduction of an extra $\mu$ destroys the Heaviside-Larmor symmetry in the predictions. As we shall see in Section~\ref{sec:decay}, this substantially changes the predictions.

We should here return to Refs.~\cite{scheelDuality, scheelDualityExtended}, as we are not the first to discuss Heaviside-Larmor symmetry in the context of macroscopic quantum electrodynamics, and its importance for decay rates and forces. Indeed, in Refs.~\cite{scheelDuality, scheelDualityExtended}, the authors show that atom-light coupling appear to break the Heaviside-Larmor symmetry of Maxwell's equations, which is predicated on the form of the coupling. The symmetry can be restored by embedding the dipole in a vacuum cavity inside the medium, thus avoiding any interaction with the noise polarisations and magnetisations. This is a type of local-field correction; a way of connecting the microscopic fields that interact with the atoms and emitters, to the macroscopic fields that are ultimately measured. We find that the vacuum-cavity model is not the whole story. 

Based partly on the arguments mentioned above, we wish to here resolve the issue by elevating the use of local fields from a `correction' to a `necessity': Heaviside-Larmor symmetry is otherwise broken. This resolves the form of the interaction energy, given that we are not interested in the internal dynamics of the magnetic dipole (as is commonly assumed also for electric dipoles), and might give insight into the force. There are however, many local field models. Indeed, in Ref.~\cite{scheelDuality}, it is argued that it is the presence of the noise polarisation $\Ph_N$ and magnetisation $\Mh_N$ that breaks Heaviside-Larmor symmetry. We will here show that this is not the case by considering another commonly used local-field model in which the dipole is allowed to interact with the noise fields. Indeed, we show that such a model still restores Heaviside-Larmor symmetry, and in the process we will consider a dual symmetric formulation of macroscopic QED that yields the same predictions but which may simplify calculations. Before this however, in the next section, let us demonstrate with an example that the form of the coupling yields qualitatively different results in magnetic media.

\section{Decay rates in absorbing media}\label{sec:decay}
Let us study the spontaneous decay rate of the magnetic dipole $\mhh$ embedded in the magneto-dielectric medium at position $\mb{r}_A$. For simplicity, we will consider a two-level emitter with transition frequency $\om_A$ coupling to the field through either $\cH^B_\tm{int} = -\mhh\cdd\Bh$ or $\cH^H_\tm{int} = -\mhh\cdd\Hh$. We will here assume that any self-interaction term $\propto \mhh^2$ is already taken into account in the internal dynamics.\footnote{This is identical to the common practise for an embedded electric dipole. Indeed, we noted earlier that the commonly used electric field operator `$\Eh$' represents the displacement field $\Eh+\dhh$, with the cavity QED-style self-interaction accounted for in the internal dynamics. In this way, we will denote `$\Hh$' as the magnetic induction $\Hh+\mhh$, equivalent to `$\Eh$' above. Whilst we recognise that this notation is a little confusing, it is nonetheless commonly adopted.} The decay rate is thus given by
\begin{align}\label{eq:decayFormalB}
\gamma_{B} = 2\pi \int_0^\infty d\om \; m_i \brakk{0}{\Bh_{i}\left(\mb{r},\om\right)\Bh^\da_{j}\left(\mb{r}_A,\om_A\right)}{0} m_j,    
\end{align}
or alternatively
\begin{align}\label{eq:decayFormalH}
\gamma_{H} = 2\pi \int_0^\infty d\om \; m_i \brakk{0}{\Hh_{i}\left(\mb{r},\om\right)\Hh^\da_{j}\left(\mb{r}_A,\om_A\right)}{0} m_j,    
\end{align}
where in both cases the limit $\mb{r}\rightarrow \mb{r}_A$ has to be taken with some care in absorbing media, and $\mb{m}$ is the magnetic transition dipole moment. Here Einstein summation convention is implied. We should note that, in principle and similarly to an electric dipole coupling to the electric field \cite{steves10}, we would have both a transverse and a longitudinal contribution to the decay rate. This is the magnetic analogue of the Joule heating via longitudinal coupling of the atom to the dielectric discussed in Ref.~\cite{steves10}. Such a contribution is absent when considering $\cH^B_\tm{int} = -\mhh\cdd\Bh$, as the magnetic induction must be purely transverse. However, we will focus on the radiative decay here and thus only consider the transverse part, as the longitudinal decay is a pure heating process which is consequently difficult to measure. Nonetheless, we note that the derivation for the longitudinal decay follows similarly to what is presented here.

\subsection{$\Hh$-like coupling}
Let us start this by evaluating $\gamma_H$ in Eq.~\eqref{eq:decayFormalH}, as this takes the simplest form in our chosen formulation. For this, we need to evaluate
\begin{align}
&\brakk{0}{\Hh_{i}(\mb{r},\om)\Hh^\da_{j}(\mb{r}_A,\om_A)}{0} \\ 
&\hspace{2cm}=\pi^{-1}\om^2 \delta(\om-\om_A)\Im\left[\gG_{ij}^{H}(\mb{r},\mb{r}_A,\om)\right],\nonumber
\end{align}
where we have left the details to Appendix~\ref{app:decayDetails} for brevity. Furthermore, when taking the limit of $\rb \rightarrow \rb_A$, we will need to introduce a spatial average $\avgg{...}$ that is required to regularise the expression, since we allow for absorption by the medium. This is simply a reflection of the macroscopic approach only being valid above the scale of the atomic separation in the medium. The spatial average is performed over a Gaussian sphere of radius $R$ \cite{steves10} such that $|n \om R| \ll 1$. Thus we have 
\begin{align}
\gamma_H = 2\omega_A^2 m^2 \Im\left[\avgg{\gG_{ij}^{H}(\mb{0},\om)}\right].
\end{align}
Prior to spatial averaging, the transverse Green's function is given most readily \cite{steves10} as 
\begin{align}
\gG^{H}_{ij}(\mb{r},\mb{r}',\om) &= \varepsilon(\om)\intkf e^{i \bk\cdd\left(\rb-\rb'\right)} \frac{\delta_{ij}-(k_i k_j/k^2)}{k^2-\om^2\varepsilon(\om)\mu(\om)} \\
&= \frac{\varepsilon(\om)}{4\pi}\left[\frac{\rho_i\rho_j}{2\rho^3}+\frac{\delta_{ij}}{2\rho}+\frac{2i}{3}\om \,n(\om)+\mathcal{O}(\rho)\right] 
%
\end{align}
where we have defined $\bs{\rho} = \mb{r}-\mb{r}'$. We now perform the spatial average over a small Gaussian sphere of radius $R$ such that $|n \om R| \ll 1$, as in
\begin{widetext}
\begin{align}
\avgg{\gG^H_{ij}(\mb{0},\om)} &\equiv \intr\intrp \left(\frac{2}{R^2}\right)^3 e^{-2\pi(r^2+r'^2)/R^2}\gG^H_{ij}(\mb{r},\mb{r}',\om) \label{eq:avgTransverse}\\
&= \varepsilon(\om)\intr\intrp\intkf e^{i\mb{k}\cdd\left(\mb{r}-\mb{r}'\right)} \left[\left(\frac{2}{R^2}\right)^3 e^{-2\pi(r^2+r'^2)/R^2}\right]\frac{\delta_{ij}-\frac{k_i k_j}{k^2}}{k^2-\om^2\mu\varepsilon}\nonumber\\
&\simeq \frac{\varepsilon(\om) \delta_{ij}}{6\pi}\left[\frac{2}{R}+i\, n(\om)\om+\mathcal{O}(R)\right],
%
\end{align}
\end{widetext}
where we in the second step used the Fourier representation of $\gG^{H}$, and expanded for $|n \om R| \ll 1$ in the final step. We thus arrive at
\begin{align}\label{eq:decayH}
\gamma_{H} &= \gamma_0\left[\Re[n\varepsilon]+\frac{2\Im[\varepsilon]}{\omega_A R}\right],
\end{align}
where $\gamma_0 = m^2 \omega_A^3/(3\pi)$ is the free-space decay rate. The last term represents a heating process, where a virtual photon is emitted and immediately absorbed by the medium. Note that Eq.~\eqref{eq:decayH} is dual symmetric under the transform considered here. Also, as is noted in Ref.~\cite{stevesLocalField}, we want $R$ to be greater than the average medium separation but smaller than the wavelength.

\begin{figure*}
    \centering
    \includegraphics[width=\textwidth]{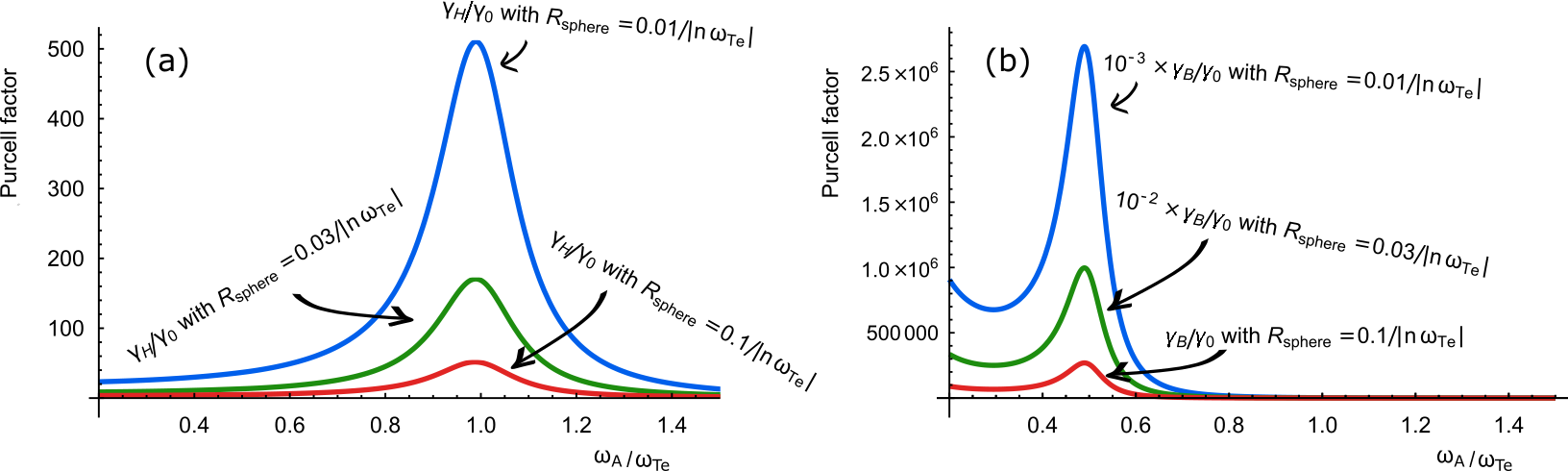}
    \caption{\textbf{(a)} Purcell factor $\gamma_H/\gamma_0$ as a function of dipole resonance $\omega_A$ in the example medium seen in Fig.~\ref{fig:disp}. \textbf{(b)} Purcell factor $\gamma_B/\gamma_0$ as a function of dipole resonance $\omega_A$ in the example medium seen in Fig.~\ref{fig:disp}. In both cases, we have chosen the spherical radius $R_\tm{sphere}$ such that $R_\tm{sphere}^3 = 3 R^3/4\pi$. With reference to an electric resonance at $\om_\tm{Te} = 2\pi c/100\tm{ nm}$, $R_\tm{sphere} \simeq \{1.27,\tm{ } 3.82,\tm{ } 12.7 \}$ {\aa}ngstr{\"o}m, respectively.}
    \label{fig:gammaHB}
\end{figure*}

The Purcell factor $\gamma_H/\gamma_0$ for the decay rate~\eqref{eq:decayH} can be found in Fig.~\ref{fig:gammaHB}(a) for an emitter at frequency $\om_A$ embedded in our example medium (with $\varepsilon$ and $\mu$ as given in Eqns.~\eqref{eq:permitt} and \eqref{eq:permea}, respectively), for $R^3 = 4\pi R^3_\tm{sphere}/3$ chosen such that $|n(\om_\tm{Te})\om_\tm{Te} R_\tm{sphere}| = \{0.01,\tm{ } 0.03,\tm{ } 0.1 \}$. If we set $\om_\tm{Te} = 2\pi c/100\tm{ nm}$ as an example, this corresponds to $R_\tm{sphere} \simeq \{1.27,\tm{ } 3.82,\tm{ } 12.7 \}$ {\aa}ngstr{\"o}m, respectively. Interestingly, we see that the Purcell factor is dominated by the near-field emission and absorption process that is $\propto \Im[\varepsilon]/(\om_A R)$, especially when the emitter frequency $\om_A$ is close to the electric resonance $\om_\tm{Te}$. This could be expected from the induced magnetic field $\Hh$ being directly proportional to the permittivity $\varepsilon$.

\subsection{$\Bh$-like coupling}
The decay rate for a magnetic induction-like coupling in Eq.~\eqref{eq:decayFormalB} follows from this result, given that we recall that in this formulation we have
\begin{align}
\Bh = \mu\Hh + \Mh_{N,H}.
\end{align}
We thus need to evaluate the correlator
\begin{align}\label{eq:correlatorBh}
&\brakk{0}{\Bh_{i}(\mb{r},\om)\Bh^\da_{j}(\mb{r'},\om')}{0} = \\
&\quad\quad \left|\mu\right|^2\brakk{0}{\Hh_{i}(\mb{r},\om)\Hh^\da_{j}(\mb{r'},\om')}{0}\nonumber\\
&\quad\quad+\brakk{0}{\Mh_{N,H,i}(\mb{r},\om)\Mh^\da_{N,H,j}(\mb{r'},\om')}{0}\nonumber\\
&\quad\quad +\mu\brakk{0}{\Hh_{i}(\mb{r},\om)\Mh^\da_{N,H,j}(\mb{r'},\om')}{0}\nonumber\\
&\quad\quad +\mu^*\brakk{0}{\Mh_{N,H,j}(\mb{r},\om)\Hh^\da_{j}(\mb{r'},\om')}{0},\nonumber
\end{align} 
and follows the same route as Ref.~\cite{stevesLocalField} with the additional complication of both $\varepsilon$ and $\mu$ being complex quantities, and some differing coefficients. We find that 
\begin{align}\label{eq:decayB}
\gamma_B = &\left|\mu\right|^2\gamma_H + 2m_i m_j \Im[\mu]\avgg{\delta_{\perp ij}(\mb{0})}  \\
&\;+4 m_i m_j\Im[\mu]\om_A^2\Re\left[\mu\avgg{\gG^H_{ij}(\mb{0},\om_A)}\right],\nonumber
\end{align}
where we've kept only the transverse parts as previously discussed. For this, we also require the spatial average
\begin{align}\label{eq:avgDelta}
\avgg{\delta_{ij}\delta(\mb{0})} &= \avgg{\delta_{\perp ij}(\mb{0})}+\avgg{\delta_{|| ij}(\mb{0})} \\
&= \frac{2\delta_{ij}}{3 R^3}+\frac{\delta_{ij}}{3 R^3} = \frac{\delta_{ij}}{R^3},\nonumber
\end{align} 
which is performed similarly to the Eq.~\eqref{eq:avgTransverse}. Finally, we find the decay rate
\begin{align}
\gamma_B = \gamma_0 \bigg[&|\mu|^2\lel(\Re[n\varepsilon]+\frac{2\Im[\varepsilon]}{\omega_A R}\rer)\\
&+\Im\lel[\mu\rer]\left( \frac{4\pi}{(\om_A R)^3} + \frac{4\Re\lel[n^2\rer]}{\omega_A R}-4\Im\lel[n^3\rer] \right)\bigg].\nonumber
\end{align}
This is quite clearly not dual symmetric. As can be seen in Fig.~\ref{fig:gammaHB}(b), the Purcell factor $\gamma_B/\gamma_0$ is qualitatively different from $\gamma_H/\gamma_0$ under the same conditions. Indeed, the Purcell factor $\gamma_B/\gamma_0$ is peaked at the magnetic resonance $\om_\tm{Tm}=\om_\tm{Te}/2$. We note that the dominating term $\propto\Im[\mu]/(\om_A R)^3$ originates from the $\brakk{0}{\Mh_{N,H,i}(\mb{r},\om)\Mh^\da_{N,H,j}(\mb{r'},\om')}{0}$-correlator: this is the near-field dipole-dipole energy transfer from the magnetic emitter to the medium.

\section{Local-field models and dual symmetric formulations}\label{sec:LL}

Let us now take local-field corrections into account. The decay rate is now given by
\begin{align}\label{eq:decayLocalFormal}
\gamma = 2\pi \int_0^\infty d\om \; m_i \brakk{0}{\Bh_{\tm{loc},i}\left(\mb{r},\om\right)\Bh^\da_{\tm{loc},j}\left(\mb{r}_A,\om_A\right)}{0} m_j.    
\end{align}
This is the magnetic analogue of Ref.~\cite{stevesLocalField}, and we can use much of the same procedure. Firstly, the local field used in Refs.~\cite{scheelDuality, scheelDualityExtended} is commonly referred to as the Onsager local-field, and assume that that the emitter is embedded in a vacuum cavity inside the medium. As mentioned, this local-field model is not unique, and we would point the reader to Ref.~\cite{polarisability} for a summary. We will here focus on the Clausius-Mossotti local-field \cite{mossotti, clausius} instead, which is just as commonly employed, especially the the magnetic media literature \cite{magnets}. This local field can be derived in two ways\footnote{Either by averaging the dipole-response surrounding the emitter or by assuming a static magnetisation in a section around the emitter \cite{LLderivations}.} based on a \textit{virtual cavity}. The virtual cavity size does not impact the result, though relies on being larger than the typical medium constituent separation, and both derivations yield 
\begin{align}
\Bh_\tm{loc} = 2\Hh/3 + \Bh/3 =\Hh_\tm{loc}. 
\end{align}
Note that $\Bh_\tm{loc} = \Hh_\tm{loc}$ is important for restoring Heaviside-Larmor symmetry to the system. This follows from the local-field models by construction, as the centre of the sphere used in their derivation is taken to be a vacuum. Furthermore, it should also be noted that the magnetic field $\Hh$ plays a much larger role than $\Bh$. If we briefly return to whether the magnetisation is a function of the magnetic field $\Hh$ or the magnetic induction $\Bh$: $\Mh(\Hh)$ or $\Mh(\Bh)$. We find that there are two options, \eqref{option1} and \eqref{option2}, as to how to include the noise magnetisation, yielding the local fields
\begin{enumerate}
    \item $\Hh_\tm{loc} = \left(\mu+2\right)\Hh/3+\Mh_{N,H}/3$,
    \item $\Hh_\tm{loc} = \left(\mu+2\right)\Hh/3+\mu\Mh_{N,B}/3$,
\end{enumerate}
respectively. Let us first consider the route of~\ref{option1}. This is indeed completely dual to the electric field $\Eh_\tm{loc}$ as considered in \citet{stevesLocalField}. The remainder of the calculation involves evaluating correlators of the form
\begin{align}\label{eq:correlator}
&\brakk{0}{\Hh_{\tm{loc},i}(\mb{r},\om)\Hh^\da_{\tm{loc},j}(\mb{r'},\om')}{0} = \\
&\quad\quad \left|\frac{\mu+2}{3}\right|^2\brakk{0}{\Hh_{i}(\mb{r},\om)\Hh^\da_{j}(\mb{r'},\om')}{0}\nonumber\\
&\quad\quad+\frac{1}{9}\brakk{0}{\Mh_{N,H,i}(\mb{r},\om)\Mh^\da_{N,H,j}(\mb{r'},\om')}{0}\nonumber\\
&\quad\quad +\left(\frac{\mu+2}{9}\right)\brakk{0}{\Hh_{i}(\mb{r},\om)\Mh^\da_{N,H,j}(\mb{r'},\om')}{0}\nonumber\\
&\quad\quad +\left(\frac{\mu^*+2}{9}\right)\brakk{0}{\Mh_{N,H,j}(\mb{r},\om)\Hh^\da_{j}(\mb{r'},\om')}{0}\nonumber\\
&= \brakk{0}{\Bh_{\tm{loc},i}(\mb{r},\om)\Bh^\da_{\tm{loc},j}(\mb{r'},\om')}{0},
\end{align} 
and follows the same route as Section~\eqref{sec:decay}.\footnote{Please see Appendix~\ref{app:decayDetails} for details.} We find that 
\begin{align}\label{eq:decayLL}
\gamma = &\left|\frac{\mu+2}{3}\right|^2\gamma_H + \frac{2m_i m_j }{9}\Im[\mu]\avgg{\delta_{\perp ij}(\mb{0})}  \\
&\;+\frac{4 m_i m_j}{3}\Im[\mu]\om_A^2\Re\left[\left(\frac{\mu+2}{3}\right)\avgg{\gG^H_{ij}(\mb{0},\om_A)}\right],\nonumber
\end{align}
where once again $\avgg{...}$ denotes a spatial average that is required to regularise the expression, and keeping the transverse part only. Also, here we denote the decay rate before local-fields are taken into account as $\gamma_H$ from Eq.~\eqref{eq:decayH}. Finally, this yields the spontaneous decay rate is given by $\gamma$ with
\begin{align}\label{eq:decayFinal}
\gamma&= \gamma_0\bigg[\left|\frac{\mu+2}{3}\right|^2\left(\Re[n\varepsilon]+\frac{2\Im[\varepsilon]}{\omega_A R}\right) +\frac{4\pi\Im[\mu]}{9(\omega_A R)^3} \nonumber\\
&\quad +\frac{4\Im[\mu]}{9}\left(\frac{\Re[n^2 + 2\varepsilon]}{\om_A R}-\Im\lel[n^3/2 + n\varepsilon\rer]\right) \bigg],
%
\end{align}
where $\gamma_0 = m^2 \omega_A^3/(3\pi)$ is the free-space decay rate. 
\begin{figure*}
    \centering
    \includegraphics[width=\textwidth]{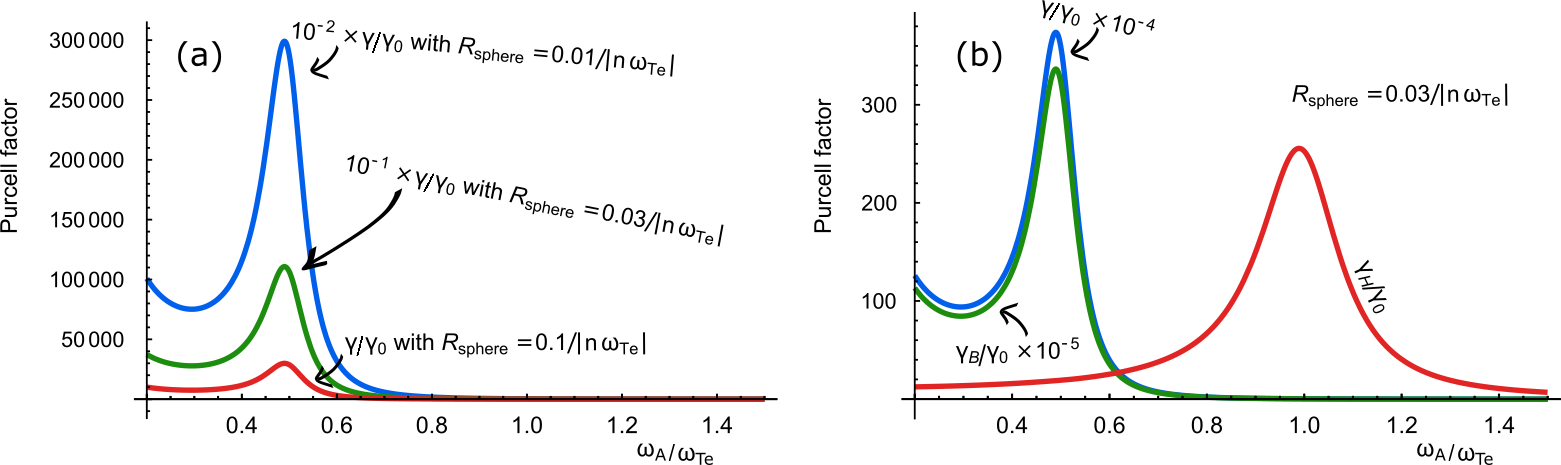}
    \caption{\textbf{(a)} Purcell factor $\gamma/\gamma_0$ as a function of dipole resonance $\omega_A$ in the example medium seen in Fig.~\ref{fig:disp}. As with Fig.~\ref{fig:gammaHB}, we have chosen the spherical radius $R_\tm{sphere}$ such that $R_\tm{sphere}^3 = 3 R^3/4\pi$, and chosen $R_\tm{sphere} \simeq \{1.27,\tm{ } 3.82,\tm{ } 12.7 \}$ {\aa}ngstr{\"o}m, respectively, with reference to an electric resonance at $\om_\tm{Te} = 2\pi c/100\tm{ nm}$. \textbf{(b)} A comparison of Purcell factors $\gamma_H/\gamma_0$, $\gamma_B/\gamma_0$ and $\gamma/\gamma_0$ as a function of dipole resonance $\omega_A$ in the example medium seen in Fig.~\ref{fig:disp}.}
    \label{fig:gammaComp}
\end{figure*}

We can find the associated Purcell factor $\gamma/\gamma_0$ in Fig.~\ref{fig:gammaComp}(a) for the same conditions as was considered in Section~\ref{sec:decay}. Note that it has a similar qualitative structure to $\gamma_B/\gamma_0$, although the peak magnitude is about an order of magnitude different. This is particularly clear in Fig.~\ref{fig:gammaComp}(b), where a comparison between the different decay rates discussed in the manuscript can be found. Indeed, the strong increase at the magnetic resonance $\om_\tm{Tm}= \om_\tm{Te}/2$ has the same origin as for $\gamma_B$, i.e. from the resonant dipole-dipole energy transfer. Importantly however, this decay rate is dual symmetric and $\mu \leftrightarrow \varepsilon$ yields the electric dipole decay rate in the same medium.

\subsection{Comparison with the option \eqref{option2}}
If we move onto option~\eqref{option2}, i.e. a formulation where $\Mh(\Bh)$ using $\Mh_{N,B}$. We can now arrive at the following decomposition
\begin{align}\label{eq:Bnoise}
\left(\begin{array}{c}
\Ph_N \\
\Mh_{N,B}
\end{array}\right) = \frac{1}{\sqrt{\pi}}\left(\begin{array}{cc}
i\sqrt{\Im{\;\varepsilon}} & 0 \\
0 & \frac{\sqrt{\Im{\;\mu}}}{|\mu|}
\end{array}\right) \left(\begin{array}{c}
\fh_e \\
\fh_m
\end{array}\right),
\end{align}
as can be found in for instance \cite{scheelUnified, scheelAnother,scheelDuality, Matloob}, amongst others. The appearance of $|\mu|$ comes as a consequence of using $\kappa=1/\mu$ in the formulation instead of $\mu$. Indeed, we usually find the fluctuation-dissipation theorem written in the form
\begin{align}\label{eq:fluctDissB}
&\brakk{0}{\Mh_{N,B,i}(\mb{r},\om)\Mh^\da_{N,B,j}(\mb{r'},\om')}{0} \nonumber\\ 
&\quad\quad = \left(\frac{-\Im[\kappa]}{\pi}\right)\delta_{ij}\delta(\mb{r}-\mb{r}')\delta(\om-\om'),
\end{align}
as $-\Im[\kappa] = -\Im[1/\mu]$ is the dissipative part of $\Bh/\mu$. We would however note that $-\Im[\kappa] = \Im[\mu]/|\mu|^2$.

As is pointed out in Ref.~\cite{scheelDuality}, this does not seem dual symmetric, and in particular the relation between the polariton operators $(\fh_{e},\fh_{m})$ and $(\Ph_N,\Mh_{N,B})$ yields an apparent dual-asymmetric relation $\fh_e^\st = -i(\mu/|\mu|)\fh_m$ and $\fh_m^\st = -i(|\varepsilon|/\varepsilon)\fh_e$. Indeed, neither $\Hh(\mb{r},\om)$\footnote{$\Hh$ transforms correctly here, but care is advised if expressed in terms of a Green's function as such a formulation is not dual symmetric.} nor $\Mh^\da_{N,B,j}(\mb{r'},\om')$ are individually dual symmetric in this formulation, following the rules outlined in Table~1 of Ref.~\cite{scheelDuality}. Regardless of the above considerations however, it can be shown that we arrive at the same decay rate $\gamma$. This is because, despite the operators being dual-asymmetric, expectation values of the form 
\begin{align}
\mu^*\brakk{0}{\Hh_{i}(\mb{r},\om)\Mh^\da_{N,B,j}(\mb{r'},\om')}{0}
\end{align}
are nonetheless dual symmetric.

The resolution of this apparent contradiction lies with the choice of phase when relating the noise operators to the polariton operators in Eq.~\eqref{eq:Bnoise}. In fact, the connections between noise operators and polariton operators are only defined up to any phase. This indifference to trivial phases has been noted as early as Ref.~\cite{scheelEarly}, but also Ref.~\cite{scheelUnified}. Indeed it is easy to see that the addition of any phase will not change Eqns.~\eqref{eq:flucDissP} and \eqref{eq:flucDissM}.
This means that we are left with an undetermined phase in Eq.~\eqref{eq:Bnoise} [as well as Eq.~\eqref{eq:Hnoise}], as this is not uniquely determined by the fluctuation-dissipation theorem. We can however use Heaviside-Larmor symmetry to fix this phase. A minor modification to Eq.~\eqref{eq:Bnoise}, such that it reads
\begin{align}
\left(\begin{array}{c}
\Ph_N \\
\Mh_{N,B}
\end{array}\right) = \frac{1}{\sqrt{\pi}}\left(\begin{array}{cc}
i\sqrt{\Im{\;\varepsilon}} & 0 \\
0 & i\frac{\sqrt{\Im{\;\mu}}}{\mu}
\end{array}\right) \left(\begin{array}{c}
\fh_e \\
\fh_m
\end{array}\right),
\end{align}
both resolves this issue, and provides a nearly structurally dual symmetric formulation of macroscopic QED. In this way, the formulation of macroscopic QED is dual symmetric already at an operator-level, rather than at the level of the expectation values. The inclusion of $\mu$ into the definition of $\Mh_{N,B}$, as we do in option~\ref{option1}, would make the formulation fully dual symmetric. We would note that this is still consistent with Ref.~\cite{scheelUnified}.

\section{Concluding remarks}\label{sec:conc}

We find that the local-field must be included in treatments of emitters embedded in macroscopic media. Any other option would break Heaviside-Larmor symmetry, which is expected to hold at these scales. Furthermore, we find that a minor modification to the magnetic section of the usual formulation of macroscopic QED makes the formulation dual symmetric already at an operator level. The impact of this is not restricted to the case we consider here, and certainly simplifies calculations.

\textit{Acknowledgements.} -- We would like to acknowledge funding from the Engineering and Physical Sciences Research Council under Grants No. EP/N509668/1 and No. EP/R513222/1, as well as The Royal Society under Grants No. RP/EA/180010 and No. RP/150122. NW wishes to acknowledge support from the Royal Commission for the Exhibition of 1851.

$\dagger$ NW and AM contributed equally to this work.


%

\appendix
\onecolumngrid
\section{Local-field corrected decay rates}\label{app:decay}

Let us start by ensuring that the quantisation used is indeed consistent with previous work, after which we will present some further details on the decay rate calculation. In the former, we will be following Ref.~\cite{dispersion}, and in the latter Ref.~\cite{stevesLocalField}. 

\subsection{Consistency of the quantisation}\label{app:decayQuant}
Starting with the quantisation, the simplest way this can be done is by directly connecting the formulation to that in Ref.~\cite{dispersion, scheelUnified}. The calculations can naturally be done independently, but with little gain. Our starting point is Maxwell's equation in frequency space:
\begin{align}
\dive{}\Dh(\rb,\om) &= 0,\label{eq:M1om}\\
\dive{}\Bh(\rb,\om) &= 0,\label{eq:M2om}\\
\curl{}\Eh(\rb,\om) &= i\om \Bh(\rb,\om),\label{eq:M3om}\\
\curl{}\Hh(\rb,\om) &= -i\om\Dh(\rb,\om),\label{eq:M4om},
\end{align}
along with the constitutive equations
\begin{align}
\Dh(\rb,\om) &= \varepsilon \Eh(\rb,\om)+\Ph_N(\rb,\om)\\
\Bh(\rb,\om) &= \mu \Hh(\rb,\om) + \Mh_{N,H}(\rb,\om).
\end{align}
It follows that the fields satisfy
\begin{align}
\curl{}\left[\curl{}\Eh\right]-\om^2\varepsilon\mu \Eh &= \om^2 \mu\Ph_N+i\om\curl{}\Mh_{N,H}, \label{eq:EfieldFull}\\
\curl{}\left[\curl{}\Hh\right]-\om^2\varepsilon\mu \Hh &= \om^2 \varepsilon\Mh_{N,H}-i\omega\curl{}\Ph_{N}.
\end{align}
As is noted in the main text, we can then relate the noise polarisation and magnetisation to the system polaritons through
\begin{align}
\Ph_N &= i\sqrt{\Im[\varepsilon]/\pi}\fh_e, \\
\Mh_{N,H} &= i\sqrt{\Im[\mu]/\pi}\fh_m, \label{eq:noiseHagain}
\end{align}
where
\begin{align}
\left[\fh_{\lambda,i}(\br,\om),\fh^\da_{\lambda',j}(\br',\om')\right]=\delta_{ij}\delta_{\lambda\lambda'}\delta(\br-\br')\delta(\om-\om').
\end{align}
Also, the time-evolution is generated through the Hamiltonian 
\begin{align}
H_\tm{field} = \sum_{\lambda=e,m}\intr \int_0^\infty d\om \; \om\,\fh^\da_\la(\br,\om)\cdd\fh_\la(\br,\om)    
\end{align}
as $i\dot{\fh}_\la = \left[\fh_\la,H_\tm{field}\right] = \om \fh_\la$. When ensuring the correctness of the quantisation procedure, we will focus on Eq.~\eqref{eq:EfieldFull}, which can be rewritten as
\begin{align}
\mu^{-1}\curl{}\left[\curl{}\Eh\right]-\om^2\varepsilon \Eh &= i\om \jh_N
\end{align}
with
\begin{align}\label{eq:noiseCurrent}
\jh_N = -i\om \Ph_N + \mu^{-1}\curl{}\Mh_{N,H},
\end{align}
from which we also find $\rhoh_N=-\dive{}\Ph_{N}$
This allows us to write 
\begin{align}
\Eh(\rb,\om) = i\om\intrp \gG^E(\rb,\rb',\om)\cdd\jh_N(\rb',\om)
\end{align}
where $\gG$ satisfies
\begin{align}
\mu^{-1}\curl{}\left[\curl{}\gG^E\right]-\om^2\varepsilon \gG^E &= \id\delta(\rb-\rb').
\end{align}
It is also convenient to use
\begin{align}
\Bh &= \cP\frac{1}{i\om}\left[\curl{}\Eh\right],\\
\Dh &= -\cP\frac{1}{i\om}\left[\curl{}\Hh\right],
\end{align}
where $\cP$ stands for the principal value part of $1/\om$. Let us now formally identify $\mu^{-1}\Mh_{N,H} = \Mh_{N,B}$ and rewrite the noise current as
\begin{align}
\jh_N = -i\om \Ph_N + \curl{}\Mh_{N,B}.
\end{align}
Subsituting this into Eq.~\eqref{eq:noiseHagain} yields
\begin{align}
\mu\Mh_{N,B} &= i\sqrt{\Im[\mu]/\pi}\fh_m.
\end{align}
We now note that this is invariant under the choice of phase \cite{scheelDuality}, and any choice of the form 
\begin{align}
\Mh_{N,B} &= \frac{i e^{i\phi}}{\mu}\sqrt{\frac{\Im[\mu]}{\pi}}\fh_m.
\end{align}
is equally correct. Here $\phi$ is any phase. If we now specify $\phi$ such that
\begin{align}
\exp\left[i\phi\right] = -i\mu/|\mu|,
\end{align}
we find that $\Mh_{N,B} = \sqrt{\Im[\mu]/(\pi |\mu|^2)} \fh_m$. This formally shows that this formulation is equivalent to the one found in Ref.~\cite{dispersion}, amongst others. It follows that 
\begin{align}
\comm{\Eh_i(\rb)}{\Bh_l(\rb')} = i\epsilon_{ijk}\partial_j \left[\delta_{k l}\delta(\rb-\rb')\right],
\end{align}
as is required by the quantisation procedure. Here $\epsilon_{ijk}$ is the Levi-Cevita symbol and $\partial_j$ denotes the derivative with respect to the $j^\tm{th}$-component of $\rb$. Also, here we define $\Eh(\rb) = \intw \left[\Eh(\rb,\om)+\Eh^\da(\rb,\om)\right]$ and likewise for the magnetic induction $\Bh(\rb)$.

\subsection{Details of the decay rate calculation}\label{app:decayDetails}

\noindent Let us consider the equation for $\gG^H$, given by Eq.~\eqref{eq:GreensH} in the main text:
\begin{align}\label{eq:Greens1}
\varepsilon^{-1}\curl{\rb}\left[\curl{\rb}\gG^H(\mb{r},\mb{r}',\om)\right]-\om^2\mu \gG^H(\mb{r},\mb{r}',\om)= \id \delta(\mb{r}-\mb{r}'),
\end{align}
and we should note that
\begin{align}\label{eq:Greens2}
\varepsilon^{-1}\curl{\rb'}\left[\curl{\rb'}\gG^H(\mb{r},\mb{r}',\om)\right]-\om^2\mu \gG^H(\mb{r},\mb{r}',\om)= \id \delta(\mb{r}-\mb{r}'),
\end{align}
is equally true. By taking the conjugate (denoted $\gG^{H *}$) of the Eq.~\eqref{eq:Greens1} and renaming some variables, we find that
\begin{align}\label{eq:GreensTranspose}
\lel(\varepsilon^*\rer)^{-1}\curl{\mb{r}'}\left[\curl{\mb{r}'}\gG^{H *}(\mb{r}',\mb{s},\om)\right]-\om^2\mu^* \gG^{H *}(\mb{r'},\mb{s},\om)= \id \delta(\mb{r}'-\mb{s}), 
\end{align}
where $\curl{\mb{r}'}$ denotes the curl with respect to coordinate $\mb{r}'$. If we multiply Eq.~\eqref{eq:Greens2} by $\gG^{H *}(\mb{r}',\mb{s})$ from the right, Eq.~\eqref{eq:GreensTranspose} by $\gG^{H}(\mb{r}',\mb{s})$ from the left, subtract the latter from the former and integrate over $\rb'$, we find that
\begin{align}\label{eq:integralRel}
\intrp \bigg(\gG^{H}&(\rb,\rb',\om)\cdd\left[\omega^2 \Im\; \mu(\om)\right]\gG^{H *}(\rb',\mb{s},\om)  \\
& +\gG^{H}(\rb,\rb',\om)\cdd\frac{\Im\;\varepsilon(\om)}{|\varepsilon(\om)|^2}\curl{\rb'}\left[\curl{\rb'}\gG^{H *}(\rb',\mb{s},\om)\right]\bigg) = \gG^{H}(\rb,\mb{s},\om)-\gG^{H *}(\rb,\mb{s},\om) \equiv \Im[\gG^{H}(\rb,\mb{s},\om)],\nonumber
\end{align}
where we have integrated by parts twice in the second term involving the curl. This derivation follows from the properties of Green's functions \cite{scheelDualityExtended}.

We now have all the tools to evaluate the correlator
\begin{align}\label{eq:correlator1}
&\brakk{0}{\Hh_{\tm{loc},i}(\mb{r},\om)\Hh^\da_{\tm{loc},j}(\mb{r'},\om')}{0} = \\
&\quad\quad \left|\frac{\mu+2}{3}\right|^2\brakk{0}{\Hh_{i}(\mb{r},\om)\Hh^\da_{j}(\mb{r'},\om')}{0}\nonumber\\
&\quad\quad+\frac{1}{9}\brakk{0}{\Mh_{N,H,i}(\mb{r},\om)\Mh^\da_{N,H,j}(\mb{r'},\om')}{0}\nonumber\\
&\quad\quad +\left(\frac{\mu+2}{9}\right)\brakk{0}{\Hh_{i}(\mb{r},\om)\Mh^\da_{N,H,j}(\mb{r'},\om')}{0}\nonumber\\
&\quad\quad +\left(\frac{\mu^*+2}{9}\right)\brakk{0}{\Mh_{N,H,j}(\mb{r},\om)\Hh^\da_{j}(\mb{r'},\om')}{0}\nonumber.
\end{align} 
We will do this term-by-term:
\begin{align}
\brakk{0}{\Hh_{i}(\mb{r},\om)\Hh^\da_{j}(\mb{r'},\om')}{0} &= \ints  \intsp \gG_{ik}^{H}(\mb{r},\mb{s},\om)\gG_{jp}^{H \da}(\mb{r}',\mb{s}',\om')\times \nonumber\\
&\hspace{-2cm}\brakk{0}{\left(\om^2 \Mh_{N,H,k}(\mb{s},\om)-i\om \varepsilon^{-1}\left[\curl{\mb{s}}\Ph_{N}(\mb{s},\om)\right]_k\right)\left(\om'^2 \Mh^\da_{N,H,p}(\mb{s}',\om')+i\om' (\varepsilon^*)^{-1}\left[\curl{\mb{s}'}\Ph^\da_{N}(\mb{s}',\om')\right]_p\right)}{0} \nonumber\\
&= \pi^{-1}\om^2\delta(\om-\om')\ints  \intsp \gG_{ik}^{H}(\mb{r},\mb{s},\om)\gG_{jp}^{H \da}(\mb{r}',\mb{s}',\om)\times \nonumber\\
&\hspace{2cm}\left(\left[\omega^2 \Im\;\mu(\om)\delta_{kp}\delta(\mb{s}-\mb{s}')\right]+\frac{\Im\;\varepsilon(\om)}{|\varepsilon(\om)|^2}\left[\epsilon_{kqr}\partial^{\mb{s}}_{q}\epsilon_{r l p}\partial^{\mb{s}}_{l}\delta(\mb{s}-\mb{s}')\right]\right)\nonumber \\
&= \pi^{-1}\om^2\delta(\om-\om')\ints \gG_{ik}^{H}(\mb{r},\mb{s},\om)\left[\omega^2 \Im\;\mu(\om)\delta_{kp}\right]\gG_{pj}^{H *}(\mb{s},\mb{r}',\om) \nonumber\\
&\hspace{2cm}+\gG_{ik}^{H}(\mb{r},\mb{s},\om)\left(\frac{\Im\;\varepsilon(\om)}{|\varepsilon(\om)|^2}\left[\epsilon_{kqr}\partial^{\mb{s}}_{q}\epsilon_{r l p}\partial^{\mb{s}}_{l}\gG_{pj}^{H *}(\mb{s},\mb{r}',\om)\right]\right)\nonumber\\
&= \pi^{-1}\om^2\delta(\om-\om')\ints \bigg[ \gG^{H}(\mb{r},\mb{s},\om)\cdd\left[\omega^2 \Im\;\mu(\om)\right]\gG^{H *}(\mb{s},\mb{r}',\om)\nonumber\\
&\hspace{2cm}+\gG^{H}(\mb{r},\mb{s},\om)\cdd\left(\frac{\Im\;\varepsilon(\om)}{|\varepsilon(\om)|^2}\curl{\mb{s}}\left[\curl{\mb{s}}\gG^{H *}(\mb{s},\mb{r}',\om)\right]\right)\bigg]_{ij}\nonumber \\
&= \pi^{-1}\om^2 \delta(\om-\om')\Im\left[\gG_{ij}^{H}(\mb{r},\mb{r}',\om)\right],
\end{align}
where $\partial^{\mb{s}}_i$ denotes the partial derivative with respect to the $i^\tm{th}$ compoment of $\mb{s}$ and $\epsilon_{ijk}$ is the Levi-Cevita symbol. We have here used the relation in Eq.~\eqref{eq:integralRel} in the final step, and by $\gG^{H\da}$ we mean the conjugate transpose. Finally, in the penultimate step, we used that $\gG^{H\da}(\rb',\mb{s},\om) = \gG^{H *}(\mb{s},\mb{r}',\om)$. From this, we find that expected result
\begin{align}
\left|\frac{\mu+2}{3}\right|^2\brakk{0}{\Hh_{i}(\mb{r},\om)\Hh^\da_{j}(\mb{r'},\om')}{0} = \left|\frac{\mu+2}{3}\right|^2\frac{\om^2}{\pi}\Im\left[\gG_{ij}^{H}(\mb{r},\mb{r}',\om)\right]\delta(\om-\om').
\end{align}
Moving to the second term, this follows directly from the fluctuation-dissipation theorem, and we find
\begin{align}
\frac{1}{9}\brakk{0}{\Mh_{N,H,i}(\mb{r},\om)\Mh^\da_{N,H,j}(\mb{r'},\om')}{0} = \left(\frac{\Im[\mu]}{9\pi}\right)\delta_{ij}\delta(\mb{r}-\mb{r}')\delta(\om-\om').
\end{align}

Finally, the last two terms in Eq.~\eqref{eq:correlator1} form a complex conjugate pair, and so it suffices to evaluate the first one. From Eq.~\eqref{eq:HinNoise} of the main text, and using the fluctuation-dissipation theorem in Eq.~\eqref{eq:flucDissM}, it is straightforward to show that
\begin{align}
\brakk{0}{\Hh_{i}(\mb{r},\om)\Mh^\da_{N,H,j}(\mb{r'},\om')}{0} = \left(\om^2/\pi\right)\Im\left[\mu(\om)\right]\delta(\om-\om')\gG_{ij}^{H}(\mb{r},\mb{r}',\om),
\end{align}
from which we find
\begin{align}
\left(\frac{\mu+2}{9}\right)\brakk{0}{\Hh_{i}(\mb{r},\om)\Mh^\da_{N,H,j}(\mb{r'},\om')}{0}\nonumber &+\left(\frac{\mu^*+2}{9}\right)\brakk{0}{\Mh_{N,H,j}(\mb{r},\om)\Hh^\da_{j}(\mb{r'},\om')}{0} \\
&\quad\quad\quad\quad\quad\quad\quad = \frac{2\om^2}{3\pi}\Im\left[\mu(\om)\right]\Re\left[\left(\frac{\mu(\om)+2}{3}\right)\gG_{ij}^{H}(\mb{r},\mb{r}',\om)\right]\delta(\om-\om').\nonumber
\end{align}
Substituting this into Eq.~\eqref{eq:decayLocalFormal} of the main text now yields Eq.~\eqref{eq:decayLL}, given that we further note the need to take a spatial average for regularisation purposes, as is also done for Eqns.~\eqref{eq:decayFormalH} and \eqref{eq:decayFormalB}. Substituting this, along with the results in Eqns.~\eqref{eq:avgTransverse} and \eqref{eq:avgDelta} into Eq.~\eqref{eq:decayLL} yields the decay rate found in Eq.~\eqref{eq:decayFinal}.

\section{Dual-asymmetric operators and dual-symmetric decay rates}\label{app:duality}

We will here show that the decay rate calculated using the conventions referred to as option (2) in the main text are still dual symmetric, even if the operators involved lack this symmetry. The simplest approach to this starts at Maxwell's equations~(1)-(4) and the duality transform table found in Ref.~\cite{scheelDuality}. For clarity, after a $\pi/2$-transform within this formalism, we find that 
\begin{align}
\Eh^\st &= \Hh, \label{eq:transE}\\
\Hh^\st &= -\Eh, \label{eq:transH}\\
\dhh^\st &= \mhh, \label{eq:transd}\\
\mhh^\st &= -\dhh, \label{eq:transm}\\
\varepsilon^\st &= \mu, \label{eq:transeps}\\
\mu^\st &= \varepsilon \label{eq:transmu}
\end{align}
which transform according to Heaviside-Larmor symmetry, as well as
\begin{align}
\Ph_{N}^\st &= \mu\Mh_{N,B}, \label{eq:transP}\\
\Mh_{N,B}^\st &= -\Ph_{N}/\varepsilon, \label{eq:transM} \\
\fh_e^\st &= -i(\mu/|\mu|)\fh_m, \label{eq:transfe}\\
\fh_m^\st &= -i(|\varepsilon|/\varepsilon)\fh_e \label{eq:transfm}
\end{align}
whose transformation is dual asymmetric. The magnetic local field in this formulation is given by
\begin{align}
\Bh_\tm{loc} = \left(\mu+2\right)\Hh/3+\mu\Mh_{N,B}/3,  
\end{align}
as noted in the main text. For the decay rate in Eq.~\eqref{eq:decayLocalFormal} of the main text, we must now calculate the local field correlator $\brakk{0}{\Bh_{\tm{loc},i}(\mb{r},\om)\Bh^\da_{\tm{loc},j}(\mb{r'},\om')}{0}$, which is slightly different to Eq.~\eqref{eq:correlator} of the main text:
\begin{align}\label{eq:correlatorB}
&\brakk{0}{\Bh_{\tm{loc},i}(\mb{r},\om)\Bh^\da_{\tm{loc},j}(\mb{r'},\om')}{0} = \\
&\quad\quad \left|\frac{\mu+2}{3}\right|^2\brakk{0}{\Hh_{i}(\mb{r},\om)\Hh^\da_{j}(\mb{r'},\om')}{0}\nonumber\\
&\quad\quad+\frac{|\mu|^2}{9}\brakk{0}{\Mh_{N,B,i}(\mb{r},\om)\Mh^\da_{N,B,j}(\mb{r'},\om')}{0}\nonumber\\
&\quad\quad +\left(\frac{\mu+2}{9}\right)\mu^*\brakk{0}{\Hh_{i}(\mb{r},\om)\Mh^\da_{N,B,j}(\mb{r'},\om')}{0}\nonumber\\
&\quad\quad +\left(\frac{\mu^*+2}{9}\right)\mu\brakk{0}{\Mh_{N,B,j}(\mb{r},\om)\Hh^\da_{j}(\mb{r'},\om')}{0}\nonumber.
\end{align} 
For clarity, we will treat this term-by-term. Whilst the first term in Eq.~\eqref{eq:correlatorB} appears identical to the corresponding term in Eq.~\eqref{eq:correlator} of the main text, it is not, because the $\Hh$-field has a different equation of motion within this formalism. Indeed, we can evaluate this using the expression for $\Hh$ and the corresponding Green's function from Ref.~\cite{scheelDuality}, where
\begin{align}
\Hh(\rb,\om) = -\intrp \left(\left[\gG_{mm}(\rb,\rb',\om)/\mu+\id\delta(\rb-\rb')\right]\cdd \Mh_{N,B}(\rb',\om)+\gG_{me}(\rb,\rb',\om)\cdd\Ph_N(\rb',\om)/\mu\right),
\end{align}
Here \begin{align}
\gG_{mm}(\rb,\rb',\om) &= \curl{}\gG(\rb,\rb',\om)\times\overset{\leftarrow}{\boldsymbol{\nabla'}},\\
\gG_{me}(\rb,\rb',\om) &= \curl{}\gG(\rb,\rb',\om)\,i\om,
\end{align}
where $\times\overset{\leftarrow}{\boldsymbol{\nabla'}}$ denote the curl from the right with respect to the primed coordinates. Finally, $\gG(\rb,\rb',\om)$ is the solution to
\begin{align}
\curl{}\left(\frac{1}{\mu}\curl{}\left[-\omega^2\varepsilon\gG(\rb,\rb',\om)\right]\right)=\id \delta(\rb-\rb')
\end{align}
However, this calculation is much simplified using Maxwell's equation~(1)-(4) and the constituent relation $\Bh = \mu\Hh + \mu \Mh_{N,B}$. We find that 
\begin{align}\label{eq:HequationB}
\Hh(\rb,\om)= \intrp \gG^H(\rb,\rb',\om)\cdd\left[\mu \om^2\Mh_{N,B}(\rb',\om)-i\om\varepsilon^{-1} \curl{\rb'}\Ph_N(\rb',\om)\right].
\end{align}
Clearly the only difference is the addition of an extra factor of $\mu$ for each $\Hh$. However, the extra $|\mu|^2$ that appears in the correlator is cancelled by the corresponding $1/|\mu|^2$ in the noise magnetisation correlator of Eq.~\eqref{eq:fluctDissB}, and we arrive at the same final result as the main text. Also, it is easy to confirm that it is dual symmetric according to the transformation rules above. Indeed, 
\begin{align}
\left[\left|\frac{\mu+2}{3}\right|^2\brakk{0}{\Hh_{i}(\mb{r},\om)\Hh^\da_{j}(\mb{r'},\om')}{0}\right]^\st = \left|\frac{\varepsilon+2}{3}\right|^2\brakk{0}{\Eh_{i}(\mb{r},\om)\Eh^\da_{j}(\mb{r'},\om')}{0}.
\end{align}
For the second term, it is easy to see that it will yield the same expectation value as the corresponding term in Eq.~\eqref{eq:correlator}, as
\begin{align}
\frac{|\mu|^2}{9}\brakk{0}{\Mh_{N,B,i}(\mb{r},\om)\Mh^\da_{N,B,j}(\mb{r'},\om')}{0} = \frac{1}{9}\brakk{0}{\Mh_{N,H,i}(\mb{r},\om)\Mh^\da_{N,H,j}(\mb{r'},\om')}{0}
\end{align}
where we used Eq.~\eqref{eq:fluctDissB} from the main text and that $-\Im[\kappa] = \Im[\mu]/|\mu|^2$. Notably, this term does not at first glance look dual symmetric, but once we take into account the dual asymmetric relation $\Mh_{N,B}^\st = -\Ph_N/\varepsilon$, then we find
\begin{align}
\left[\frac{|\mu|^2}{9}\brakk{0}{\Mh_{N,B,i}(\mb{r},\om)\Mh^\da_{N,B,j}(\mb{r'},\om')}{0}\right]^\st = \frac{1}{9}\brakk{0}{\Ph_{N,i}(\mb{r},\om)\Ph^\da_{N,j}(\mb{r'},\om')}{0}.
\end{align}
The final two terms are complex conjugates, so it suffices to discuss
\begin{align}
\left(\frac{\mu+2}{9}\right)\mu^*\brakk{0}{\Hh_{i}(\mb{r},\om)\Mh^\da_{N,B,j}(\mb{r'},\om')}{0}.
\end{align}
Here we must be careful, as this takes a distinctly different from as compared to the electric calculation seen in Ref.~\cite{stevesLocalField}. Nonetheless, if we substitute $\Hh$ from Eq.~\eqref{eq:HequationB} and use Eq.~\eqref{eq:fluctDissB} of the main text, we arrive at the same result as the main text. In fact, a duality transform of Eqns.~\eqref{eq:transE}-\eqref{eq:transmu}, along with the implied dual-asymmetric transforms in Eqns.~\eqref{eq:transP}-\eqref{eq:transfm} shows that the total expectation value is indeed still dual symmetric: 
\begin{align}
\left[\left(\frac{\mu+2}{9}\right)\mu^*\brakk{0}{\Hh_{i}(\mb{r},\om)\Mh^\da_{N,B,j}(\mb{r'},\om')}{0}\right]^\st = \left(\frac{\varepsilon+2}{9}\right)\brakk{0}{\Eh_{i}(\mb{r},\om)\Ph^\da_{N,j}(\mb{r'},\om')}{0}.
\end{align}
This is because the dual symmetry is restored by the dual asymmetric relation in Eq.~\eqref{eq:transM}.

\end{document}